\documentclass[journal=jpclcd,manuscript=letter]{achemso}
\usepackage{amsmath,amssymb}
\usepackage{float}

\usepackage[T1]{fontenc}
\usepackage[utf8]{luainputenc}
\usepackage{float}
\usepackage{multirow}
\usepackage{lipsum}
\usepackage{mathrsfs}
\usepackage{setspace}
\usepackage{droidsans}
\usepackage{balance}
\usepackage{times,mathptmx}
\usepackage{braket}
\usepackage{sectsty}
\usepackage{graphicx}
\usepackage{amsmath,amssymb}
\usepackage{mathrsfs}
\usepackage{multirow} 
\usepackage{lastpage}
\usepackage{adjustbox}
\usepackage{upgreek}
\usepackage{color}
\usepackage{array}
\usepackage{float}
\usepackage{fancyhdr}
\usepackage{fnpos}
\usepackage{adjustbox}
\usepackage{charter}
\usepackage[format=plain,justification=justified,singlelinecheck=false,font={stretch=1.125,small,sf},labelfont=bf,labelsep=space]{caption}
\usepackage[english]{babel}
\DeclareUnicodeCharacter{0308}{~}
\usepackage[version=3]{mhchem} 

\author{Deepika Gill$^{\$}$}
\email{Deepika@physics.iitd.ac.in[DG]}
\author{Gunjana Yadav$^{\$}$}
\author{Saswata Bhattacharya}
\email{saswata@physics.iitd.ac.in[SB]}
\phone{+91-11-2659 1359}
\affiliation[Indian Institute of Technology Delhi]
{Department of Physics, Indian Institute of Technology Delhi, New Delhi, India}
\footnotetext{\textit{$^{\$}$ Equally contributed}}
\title[An \textsf{achemso} demo]
{Sn/Ge substitution in  ((C$_\textrm{n}$H$_{2\textrm{n}-1}$NH$_3$)$_2$PbI$_4$; n=3): An emerging 2D layered hybrid perovskites with enhanced optoelectronic properties$^\dag$}
\begin{document}

\begin{abstract}
Two-dimensional (2D) perovskites show higher stability in comparison to their three-dimensional (3D) counterparts. Therefore, 2D perovskites have invoked remarkable attention in basic understanding of their physical properties and optoelectronic applications. Here  we present a low-dimensional naturally self-assembled inorganic-organic (IO) hybrid systems based on primary  cyclic  ammonium-based (C$_{\textrm{n}}$H$_{2\textrm{n}-1}$NH$_{3}$) semiconductor series [viz. ((C$_{\textrm{n}}$H$_{2\textrm{n}-1}$ NH$_3$)$_2$PbI$_4$; n=3-6)]. However, the wide bandgap nature and presence of toxicity due to lead (Pb) prohibit their applications. Therefore, in the present work, we study the role of Ge/Sn substitution and Pb-vacancy (Pb-$\boxtimes$) to reduce concentration of Pb and to enhance solar cell efficiency by the formation of mixed perovskite structures. We have discussed the effect of spin-orbit coupling (SOC) using state-of-the-art hybrid density functional theory (DFT). We find the mixed conformers with Pb-$\boxtimes$ do not possess structural stability. Moreover, they have indirect bandgap, which is not good for solar cell applications. Only those conformers, which have favourable thermodynamics and structural stability, are considered for further study of optical properties. Our results infer that Sn substitution is more favorable than that of Ge in replacing Pb and enhancing the efficiency. Exciton binding energies calculated using Wannier-Mott approach for pristine and substituted conformers are larger than lead halide perovskites, while the electron-phonon coupling is smaller in the former. From computed spectroscopic limited maximum efficiency (SLME), these 2D perovskites show enough promise as alternatives to conventional lead halide perovskites.	
  \begin{tocentry}
  \begin{figure}[H]
  	\includegraphics[width=1.0\columnwidth,clip]{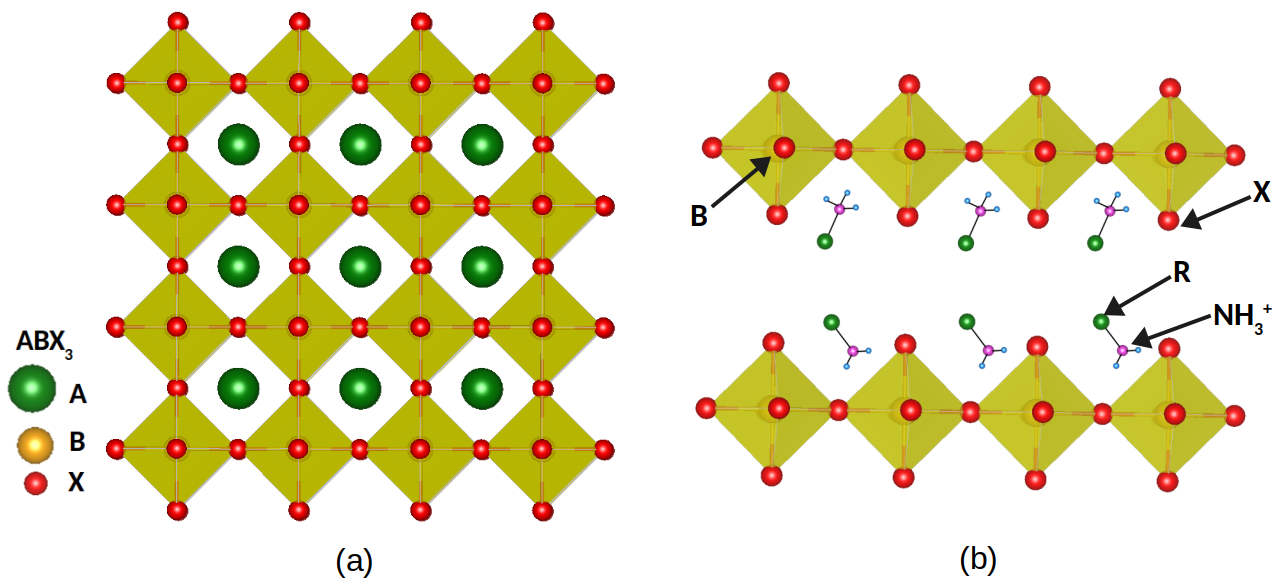}
    \end{figure}	
   \end{tocentry}
\end{abstract}
\section{Introduction}
In the past few years, 3D inorganic-organic hybrid perovskites (IOHPs) have brought revolution in the field of optoelectronics owing to their exotic optoelectronic properties.  These materials exhibit high absorption coefficient, tunable bandgap, high carrier mobility and large carrier diffusion length~\cite{snaith2013perovskites, noh2013chemical, stenberg2017perovskite, yin2014unique, kojima2009organometal, lee2012efficient, green2014emergence, gratzel2014light,  basera2020reducing, gill2021high, D0TC01484B}. Despite the huge success, poor stability (i.e. the solar cell loses efficiency during operation) and lead-toxicity have hindered their large scale commercialization~\cite{schileo2021lead, chiarella2008combined, C9TC01967G}. Thus, a sustainable future calls for the development of an efficient, cost-effective, non/less-toxic, eco-friendly and environmentally stable solar material to meet the necessity of potential energy at large scale. 

\begin{figure}[h]
 \centering
 \includegraphics[width=0.4\textwidth]{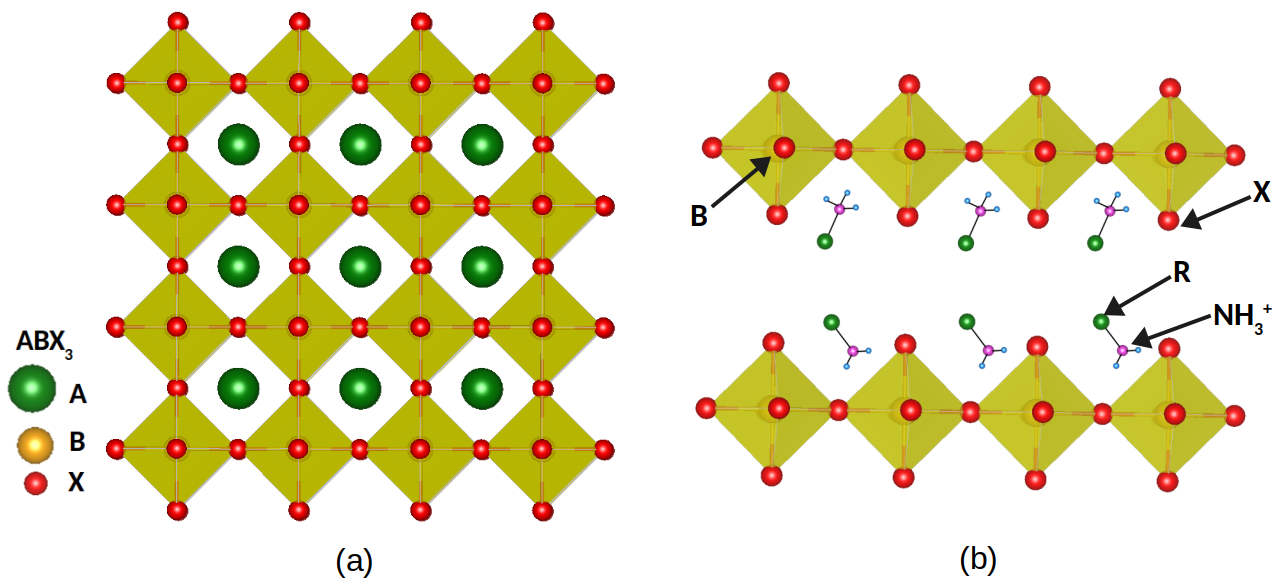}
 \caption{{\bf (a)} A 3D cubic perovskite structure with chemical formula ABX$_3$, where A, B and X are organic cation, divalent metal cation and monovalent halide anion, respectively.  {\bf (b)} Schematic drawing of 2D layered structure of the (R-NH$_3$)$_2$BX$_4$ hybrids.}
 \label{fig1}
\end{figure}

In this quest, researchers are looking into 2D layered perovskites~\cite{r1,r2,r3,r4,r5,r6}. A perfect 2D layered perovskite has the general formula (R-NH$_{3}$)$_{2}$BX$_{4}$, where R is the organic moiety, which can be derived from basic ABX$_{3}$ type perovskite structure~\cite{billing2007inorganic}. Note that in 3D perovskite, the A-site cation sits in voids of the 3D network, which have limited allowed space for A-site cations (see Fig. \ref{fig1}(a)). In 1926, Goldschmidt derived a tolerance factor ($\emph t$) formula (Equation \ref{eq:1})\cite{goldschmidt1926gesetze} that determines this allowed space i.e., maximum allowed ionic radius for A-site cation. 
For a perfect cubic perovskite structure ABX$_{3}$,
\begin{equation}
r_{\textrm{A}} + r_{\textrm{X}} = t\sqrt{2}(r_{\textrm{B}} + r_{\textrm{X}})
 \label{eq:1}
\end{equation} 
where, $\emph r_\textrm{A}$, $\emph r_\textrm{B}$, and $\emph r_\textrm{X}$ are the effective ionic radii of A-site, B-site and X-site ions, respectively. The Goldschmidt tolerance factor must be in the range 0.8 $\le$ \emph t $\le$ 1.0 for a cubic perovskite structure.\cite{wang2021lead, liang2004criteria} If B-site cation is a metal ion Pb${^{2+}}$ with $\emph r$$_\textrm{Pb}$ = 1.03 \AA, and X-site anion is a halide ion I${^-}$ with $\emph r$$_\textrm{I}$ = 2.20 \AA, then with maximum possible scenario i.e., $\emph t$ = 1.0, the geometric limit applied on A-site cation will be $\emph r$$_\textrm{A}$ = 2.36 \AA. Hence, for $\emph r$$_\textrm{A}$ > 2.36 \AA, the 3D network will be destroyed and could form 2D perovskite (see Fig. \ref{fig1}(b)). Several studies have been done in 2D perovskite structures, which showed that the 2D perovskite has more structural and chemical flexibility in comparison to their 3D counterparts. Also, the long chain organic spacers which are hydrophobic in nature of 2D perovskite can enhance the poor stability of 3D IOHPs~\cite{zhang2020revealing, ge2020recent, gill2021exploring}. However, decreasing dimensionality of IOHPs from 3D to 2D structure causes an increase in bandgap and exciton binding energy. Due to the wide bandgap nature, 2D IOHPs show poor optical absorption in PV applications\cite{traore2018composite, zhou2019two, chakraborty2021dielectric, pedesseau2016advances, boix2015perovskite, saidaminov2017low, hu2018two, ortiz2019two, ge2020recent}.

Therefore, there is justified interest to search for a stable and efficient 2D (layered) perovskite material with good optical absorption. Incidentally, we have studied and experimentally synthesized the primary cyclic ammonium-based (C$_\textrm{n}$H$_\textrm{2n-1}$NH$_\textrm{3}$; n = 3$-$6) inorganic-organic hybrid semiconductor series~\cite{dehury2021structure}. However, theoretically this system ((C$_\textrm{n}$ H$_{2\textrm{n}-1}$NH$_3$)$_2$PbI$_4$; n=3-6) is rather unexplored and requires further attention to get more atomistic insights and electronic structures. Moreover, the wide bandgap nature and presence of toxicity due to lead (Pb) prohibit their applications. Therefore, in the present work, we study the role of Ge/Sn substitution and Pb-vacancy (Pb-$\boxtimes$)  to reduce concentration of Pb and enhance solar cell efficiency by the formation of mixed perovskite structures. To do that, we have first thoroughly benchmarked and validated the exchange and correlation ($\epsilon_{xc}$) functionals in the framework of Density Functional Theory (DFT) so that the results are not any artefacts of the same. After that, we have investigated the thermodynamic stability~\cite{abjpcl} by calculating the formation energy, and structural stability~\cite{gill2021high, doi:10.1063/5.0031336} with the help of Goldschmidt tolerance factor and octahedral factor. Thereafter, we have analyzed the electronic and optical properties of the stable configurations. Finally, we have computed exciton binding energy, strength of electron-phonon coupling and the spectroscopic limited maximum efficiency (SLME) to address their suitability and theoretical maximum efficiency as a potential solar cell materials.

\begin{figure*}[hbt]
\centering
\includegraphics[width=0.5\textwidth]{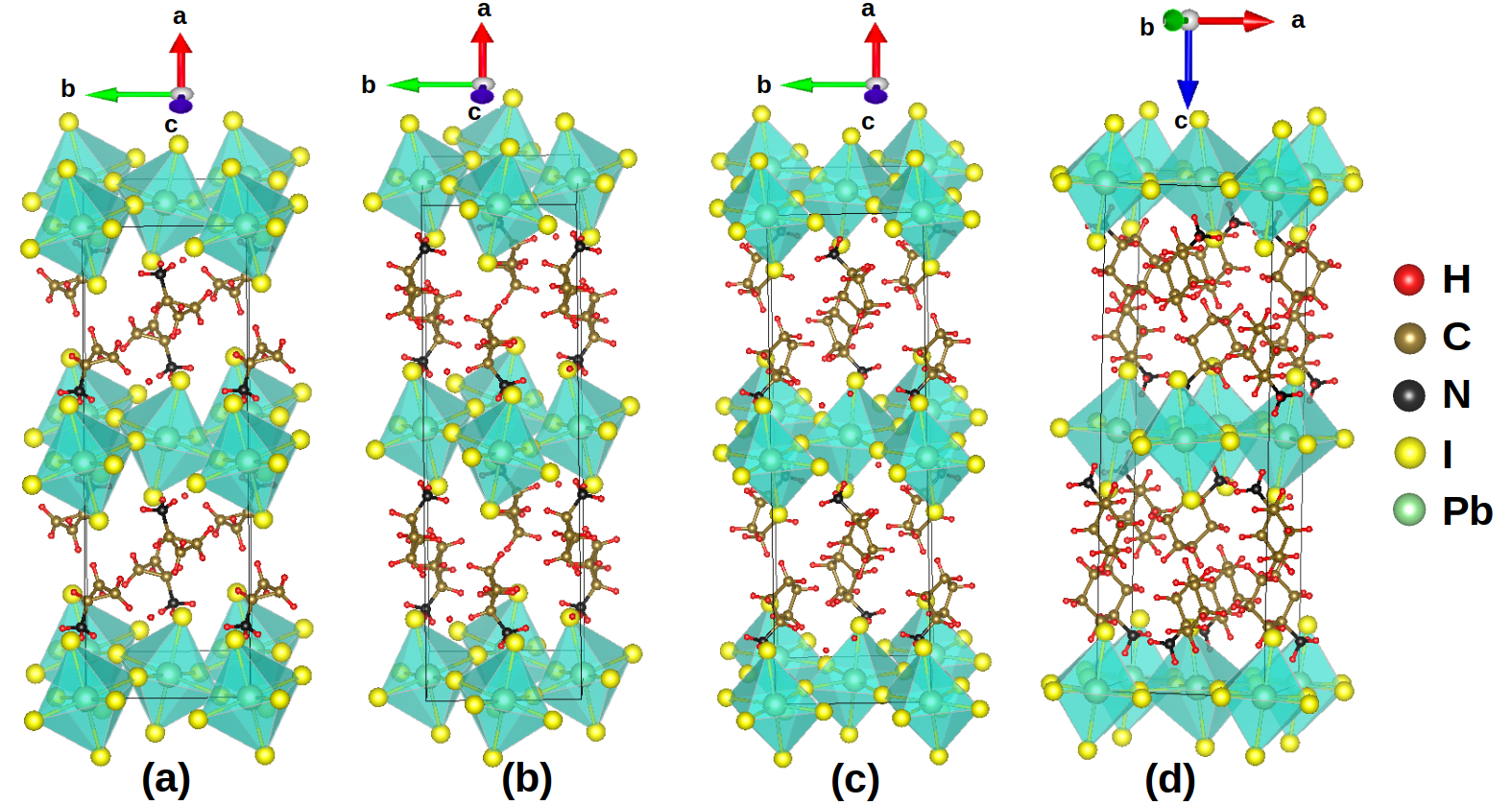}
\caption{Crystal structures for compounds: {\bf (a)} cyclopropyl ammonium tetraiodoplumbate (CPPI, n = 3), (C$_{3}$H$_{5}$NH$_{3}$)$_{2}$PbI$_{4}$, {\bf (b)} cyclobutyl ammonium tetraiodoplumbate (CBPI, n = 4), (C$_{4}$H$_{7}$NH$_{3}$)$_{2}$PbI$_{4}$, {\bf (c)} cyclopentyl ammonium tetraiodoplumbate (CPEPI, n = 5), (C$_{5}$H$_{9}$NH$_{3}$)$_{2}$PbI$_{4}$, and {\bf (d)} cyclohexyl ammonium tetraiodoplumbate (CHXPI, n = 6), (C$_{6}$H$_{11}$NH$_{3}$)$_{2}$PbI$_{4}$.}
\label{fig2}
\end{figure*}
\section{Computational Methodology}
We have performed all the calculations using Vienna $\emph{Ab initio}$ Simulation Package (VASP)\cite{kresse1996efficiency} and projector augmented-wave (PAW)\cite{blochl1994projector} pseudopotentials within the framework of DFT.\cite{hohenberg1964inhomogeneous}\cite{kohn1965self} We have optimized the crystal structures of all conformers using Perdew–Burke–Ernzerhof (PBE)\cite{perdew1996generalized} exchange-correlation ($\epsilon_{xc}$) functional with a $\Gamma$-centered 2$\times$2$\times$2 $\emph k$-point mesh, and set the criteria for convergence of total energy and forces (for optimization of atomic positions and lattice vectors) to 10${^{-5}}$ eV and 10${^{-4}}$ eV/\AA, respectively. The energy cutoff was set to 600 eV. Later on, from convergence test, we have found that a $\Gamma$-centered 3$\times$3$\times$3 $\emph k$-point mesh is sufficient for sampling the Brillouin zone (BZ), and so, the 3$\times$3$\times$3 $\emph k$-point mesh has been used in our further calculations. We have used advanced hybrid $\epsilon_{xc}$ functional Heyd–Scuseria–Ernzerhof (HSE06)\cite{heyd2003hybrid} to get more accuracy in our results because PBE functional commonly underestimates the bandgap of the materials. The spin orbit coupling (SOC) effect has been duly included in all the calculations. 
\section{Results and Discussions}
The cyclic compounds cyclopropyl ammonium tetraiodoplumbate (CPPI), cyclobutyl ammonium tetraiodoplumbate (CBPI), cyclopentyl ammonium tetraiodoplumbate (CPEPI) and cyclohexyl ammonium
tetraiodoplumbate (CHXPI) have well-defined 2D layers. There are changes in tilting of PbI$_{6}$ octahedra within layers and packing of ammonium cations between layers of these compounds, but the overall structure remains the same, i.e., 2D layered perovskite crystal structure (see Fig. \ref{fig2}).\cite{billing2007inorganic} These cyclic inorganic-organic hybrid compounds have been synthesized experimentally, and show a decrement in electronic bandgap value from n = 3 to 6, an intense narrow exciton emission, and a strong room-temperature photoluminescence.\cite{pradeesh2013synthesis, dehury2021structure} However, these compounds have some drawbacks, such as wide bandgap and presence of toxic element Pb. Therefore, to overcome these issues, which are not good for solar cell, we have studied the effect of Ge/Sn substitution and/or Pb-$\boxtimes$  using hybrid DFT. All these mentioned layered structures will show quite similar optoelectronic properties due to their similarity in crystal structures. Therefore, in present work, we have chosen one of these compounds, viz. CPPI, as our prototypical model system, and the rest of our calculations are done by considering this system.  
\subsection{Benchmarking of DFT functionals}
\begin{figure}
\centering
\includegraphics[width=0.5\textwidth]{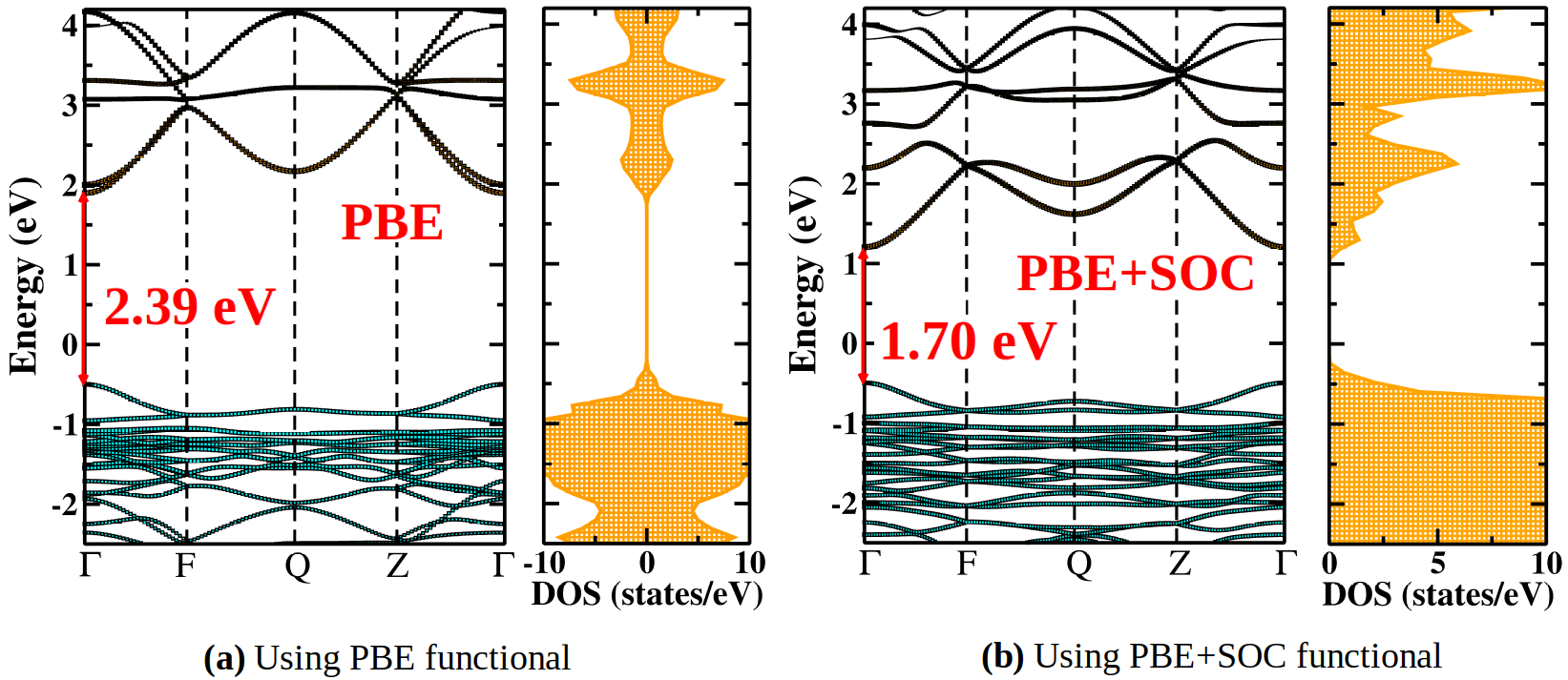}
\vspace*{0.4cm}
\includegraphics[width=0.5\textwidth]{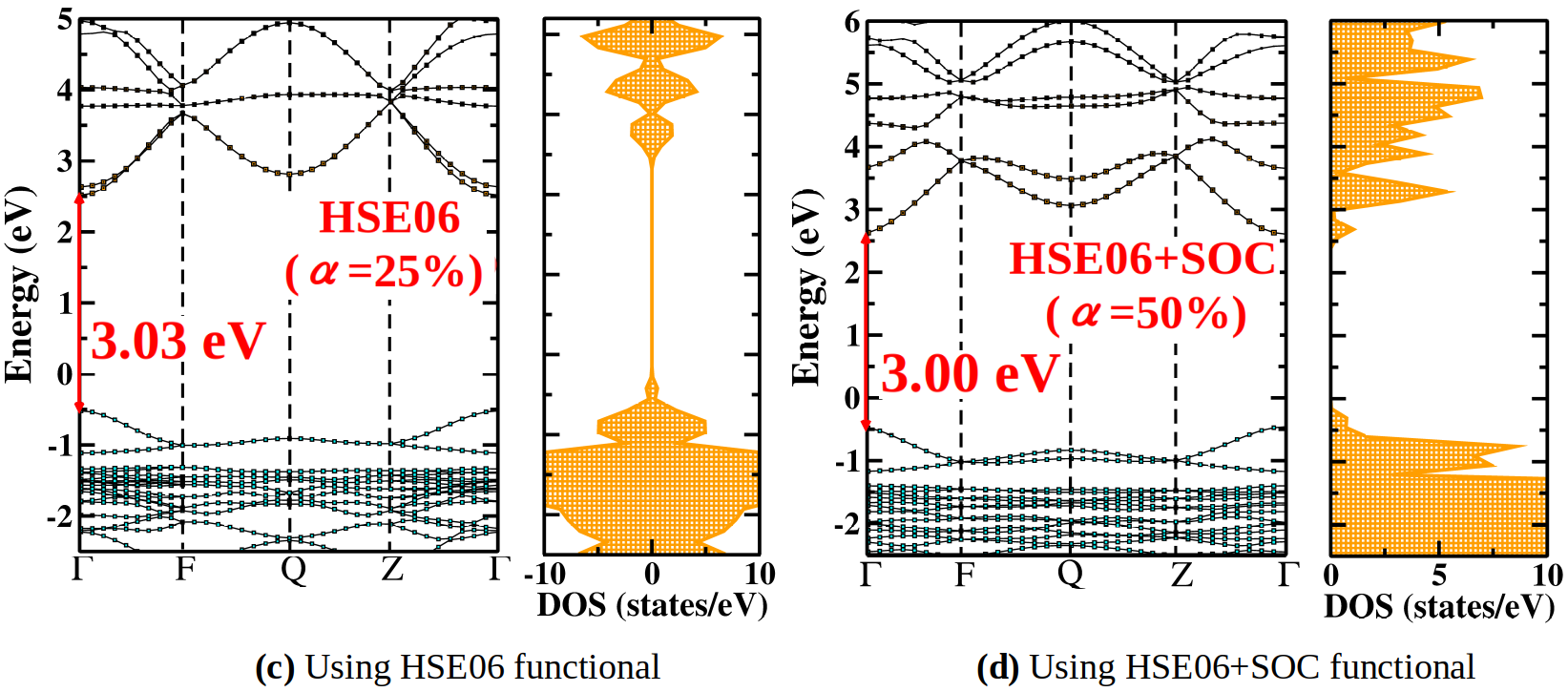}
\caption{Calculated band structures along with the density of states (DOS) of CPPI. The band paths are along the high symmetry $\emph k$-points $\Gamma$ (0, 0, 0), F (0, 0.5, 0), Q (0, 0.5, 0.5), and Z (0, 0, 0.5) of BZ.}
\label{fig3}
\end{figure}

\begin{figure}
\centering
\includegraphics[width=0.5\textwidth]{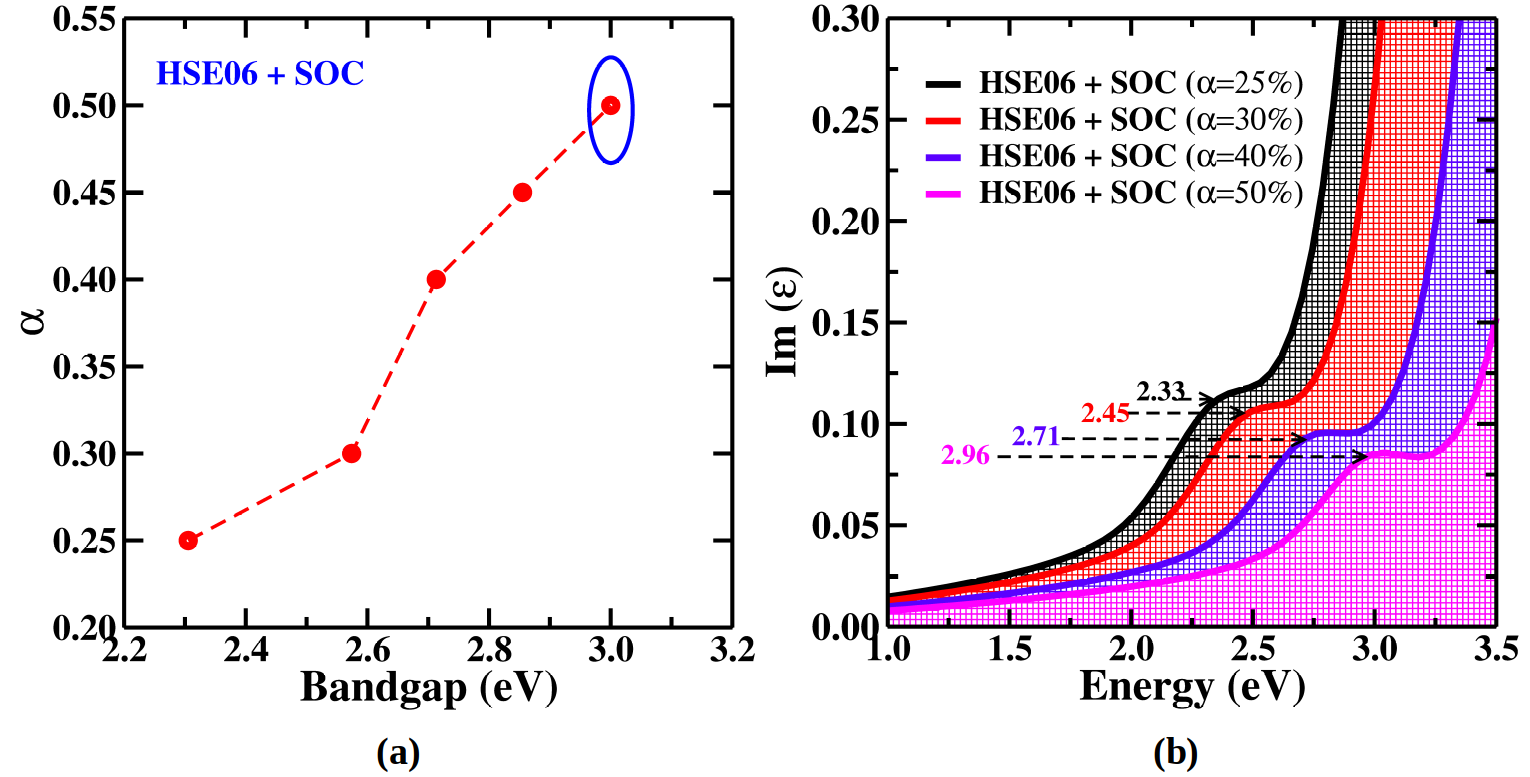}
\caption{{\bf (a)} Variation in the bandgap of CPPI with $\alpha$. The point inside blue ellipse represents the calculated bandgap, which is very close to the experimental bandgap. {\bf (b)} Imaginary part of the dielectric function calculated using HSE06+SOC with
different Hartree-Fock exact exchange (i.e., $\alpha$ = 0.25, 0.30, 0.40 and 0.50).}
\label{fig4}
\end{figure}

To ensure that our results are not merely the artefacts of DFT $\epsilon_{xc}$ functionals, we have benchmarked different $\epsilon_{xc}$ functionals by comparing the calculated bandgap (E$_\textrm{g}^\textrm{cal}$) and experimental bandgap (E$_\textrm{g}^\textrm{exp}$) of CPPI. The value of E$_\textrm{g}^\textrm{exp}$ = 3.04 eV.\cite{pradeesh2013synthesis}\cite{dehury2021structure} Using PBE functional, we have found that the value of E$_\textrm{g}^\textrm{cal}$ for CPPI is 2.39 eV (see Fig. \ref{fig3}(a)), which shows that PBE functional underestimates the E$_\textrm{g}^\textrm{exp}$ value. Since CPPI contains a heavy element Pb, we have included SOC effect with PBE functional, which results in the splitting of conduction band and the conduction band minimum (CBm) shifts to a lower value (see Fig. \ref{fig3}(b)). As a result, E$_\textrm{g}^\textrm{cal}$ comes out to be 1.70 eV using PBE+SOC. Thus, PBE functional can not reproduce the E$_\textrm{g}^\textrm{exp}$ value. After that, we have checked bandgap using HSE06 functional, which corrects the electron self-interaction error. Although HSE06 functional with default $\alpha$ = 25\% (fraction of Hartree-Fock exact exchange) is reproducing the experimental bandgap (E$_\textrm{g}^\textrm{cal}$ = 3.03 eV, see Fig. \ref{fig3}(c)) without including SOC, we need to include SOC effect due to the presence of heavy element Pb, as discussed earlier. Thereafter, we have obtained E$_\textrm{g}^\textrm{cal}$ = 2.30 eV using HSE06+SOC with $\alpha$ = 25\% (see Fig. \ref{fig4}(a)), which is also not in good agreement with the E$_\textrm{g}^\textrm{exp}$ value. Therefore, to reproduce the experimental bandgap by using HSE06+SOC, we have increased the amount of $\alpha$, which further shifts the valence band maximum (VBM) with slight alteration of CBm.
Thus, we have reproduced the E$_\textrm{g}^\textrm{exp}$ value using HSE06+SOC functional with increased amount of $\alpha$ = 50\% (see Fig. \ref{fig3}(d)). Fig. \ref{fig3} clearly depicts that band profile remains the same by both the functionals PBE and HSE06, the only difference is in the value of the direct bandgap at $\Gamma$ point.

To validate the calculations done by HSE06+SOC with different amounts of $\alpha$ (see Fig. \ref{fig4}(a)), we have calculated imaginary part of the dielectric function with four different values of $\alpha$ (i.e., $\alpha$ = 25\%, 30\%, 40\%, and 50\%) and found that the respective optical peaks are observed at 2.33, 2.45, 2.71 and 2.96 eV (see Fig. \ref{fig4}(b)). The optical peak corresponding to $\alpha$ = 50\% at 2.96 eV has a good agreement with E$_\textrm{g}^\textrm{exp}$ value (see Fig. \ref{fig4}(b)). Therefore, we have used HSE06 functional rather than PBE functional with SOC effect to achieve more accuracy in our results. Note that we have chosen the alternatives Ge, Sn and/or Pb-$\boxtimes$  to remove/reduce the amount of toxic Pb from CPPI (beacuse Ge, Sn and Pb belongs to the same group in the periodic table and have similar valence electronic configurations). For substitution of Ge/Sn to replace Pb, we have used a 58 atoms supercell, i.e., (\textrm{C}$_{3}$\textrm{H}$_{5}$\textrm{NH}$_{3}$)$_{4}$\textrm{Pb}$_{2}$\textrm{I}$_{8}$, where the Pb-defect is localized.
\subsection{Thermodynamic Stability}
\begin{figure}
\centering
\includegraphics[width=0.6\textwidth]{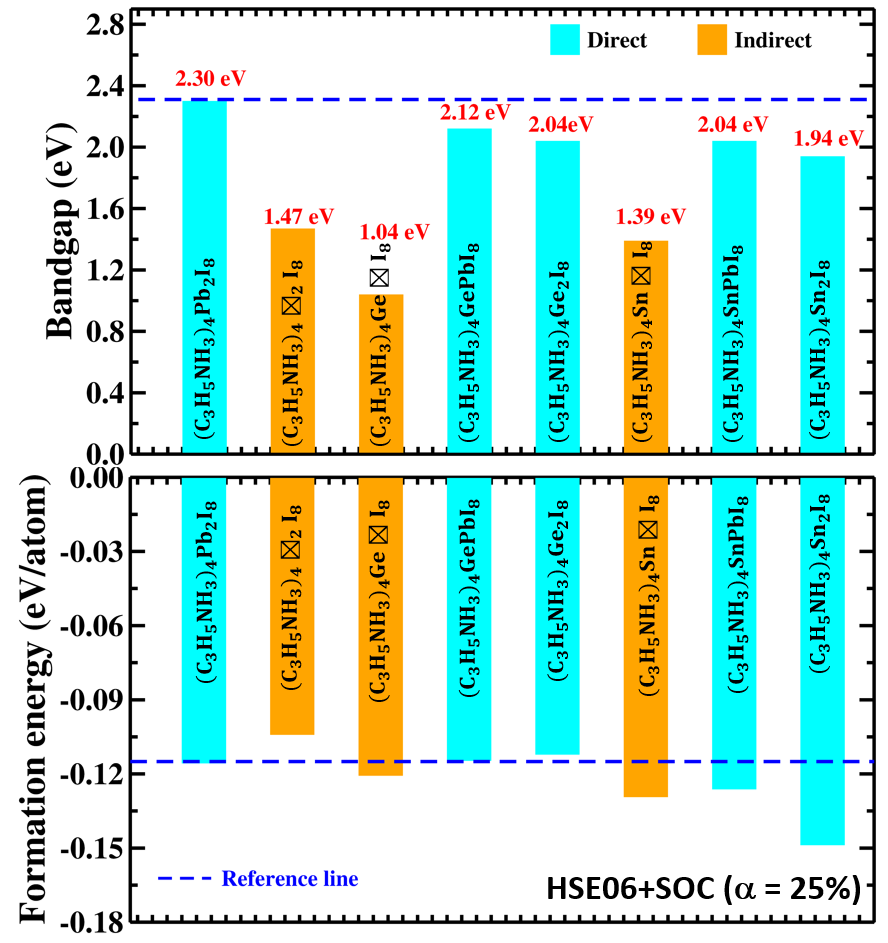}
\caption{Formation energy (eV/atom) of different mixed conformers (the blue dotted line is the reference line corresponding to prototypical material (CPPI)), and their respective bandgap using HSE06+SOC with $\alpha$ = 25\%.}
\label{fig5}
\end{figure}

Note that the amount of substitution of Ge, Sn and Pb-$\boxtimes$ will affect the SOC role in mixed perovskites, (C$_{3}$H$_{5}$NH$_{3}$)$_{2}$Pb$_{1-x-y}$Sn$_{x}$$\boxtimes$$_{y}$I$_{4}$ and (C$_{3}$H$_{5}$NH$_{3}$)$_{2}$Pb$_{1-x-y}$Ge$_{x}$$\boxtimes$$_{y}$I$_{4}$ ($\emph x$ and $\emph y$ indicates the contents of Sn/Ge and Pb-$\boxtimes$, respectively), because SOC is mainly a function of extent of Pb in this system. Thus, although the correct positions of the VBM and CBm are obtained by using HSE06+SOC with $\alpha$ = 50\% in the case of CPPI, it will not necessarily the same in the case of mixed perovskites. Hence, we have used the default value $\alpha$ = 25\% for the energy calculations of mixed conformers. The bandgaps of different conformers with the default $\alpha$ are shown in the upper panel of Fig \ref{fig5}. Note that, this may lead to some error in the total energy expression due to under/overestimation of the combined effect of SOC and the electron’s self-interaction error~\cite{basera2020reducing}. In order to eliminate this type of error, we took the difference of total energies for the calculation of the formation energy of different conformers with and without defects. In the case of Ge and Sn doping, the considered precursors 
are PbI$_{2}$, GeI$_{2}$, I$_{2}$, HI, C$_{3}$H$_{5}$NH$_{3}$ and SnI$_{2}$. We have calculated the formation energy as follows
\begin{equation}
\begin{split}
E_\textrm{f}(\emph x, \emph y)=E(\textrm{C}_{12}\textrm{H}_{32}\textrm{N}_4\textrm{Pb}_{2-\emph x - \emph y}\textrm{Ge}_{\emph x}\boxtimes_{\emph y}\textrm{I}_8)- (2-\emph x - \emph y)E(\textrm{PbI}_2)\\- \emph x E(\textrm{GeI}_2)- \emph y E(\textrm{I}_2) - 4E(\textrm{HI}) - 4E(\textrm{C}_3\textrm{H}_5\textrm{NH}_2)
\end{split}
\label{eq2}
\end{equation}
where, 0 $\le$ \emph x $\le$ 2 and 0 $\le$ \emph y $\le$ 2 in the supercell of (C$_{3}$H$_{5}$NH$_{3}$)$_{2}$PbI$_{4}$ i.e., (C$_{3}$H$_{5}$NH$_{3}$)$_{4}$Pb$_{2}$I$_{8}$. In the case of Sn substitution, SnI$_{2}$ is used instead of GeI$_{2}$ in Equation \ref{eq2}.

First, we have determined the most favourable Pb site for Ge / Sn substitution alongside existence of Pb-$\boxtimes$ via an iterative procedure~\cite{basera2020reducing}. Here, it should be noted that both Pb sites are equivalent sites. Therefore, we can substitute alternative Ge/Sn at any Pb sites. The thermodynamic stability of different mixed conformers by using HSE06 functional is given in the ESI† (see Table S1 and Fig. S1) as a reference data set to understand the explicit role of SOC effect. We have found that the mixed conformer, which has complete Pb-$\boxtimes$ with
no Ge/Sn substitution is thermodynamically unstable with respect to the CPPI (see bottom panel of Fig. \ref{fig5}, and Fig. S1). As the content of Sn substitution increases without Pb-$\boxtimes$, a gradual increase in thermodynamic stability is observed (see bottom panel of Fig. \ref{fig5}). However, as the content of Ge 
substitution increases, it shows less thermodynamic stability with respect to CPPI (see bottom panel of Fig. \ref{fig5}). Complete Sn substitution is thermodynamically the most stable one. Thus, Sn substitution is thermodynamically more preferable than Ge substitution. The crystal structures of the Sn substituted CPPI are shown in Fig. S2. 
\subsection{Structural Stability}
\begin{figure}[b]
\centering
  \includegraphics[width=0.5\linewidth]{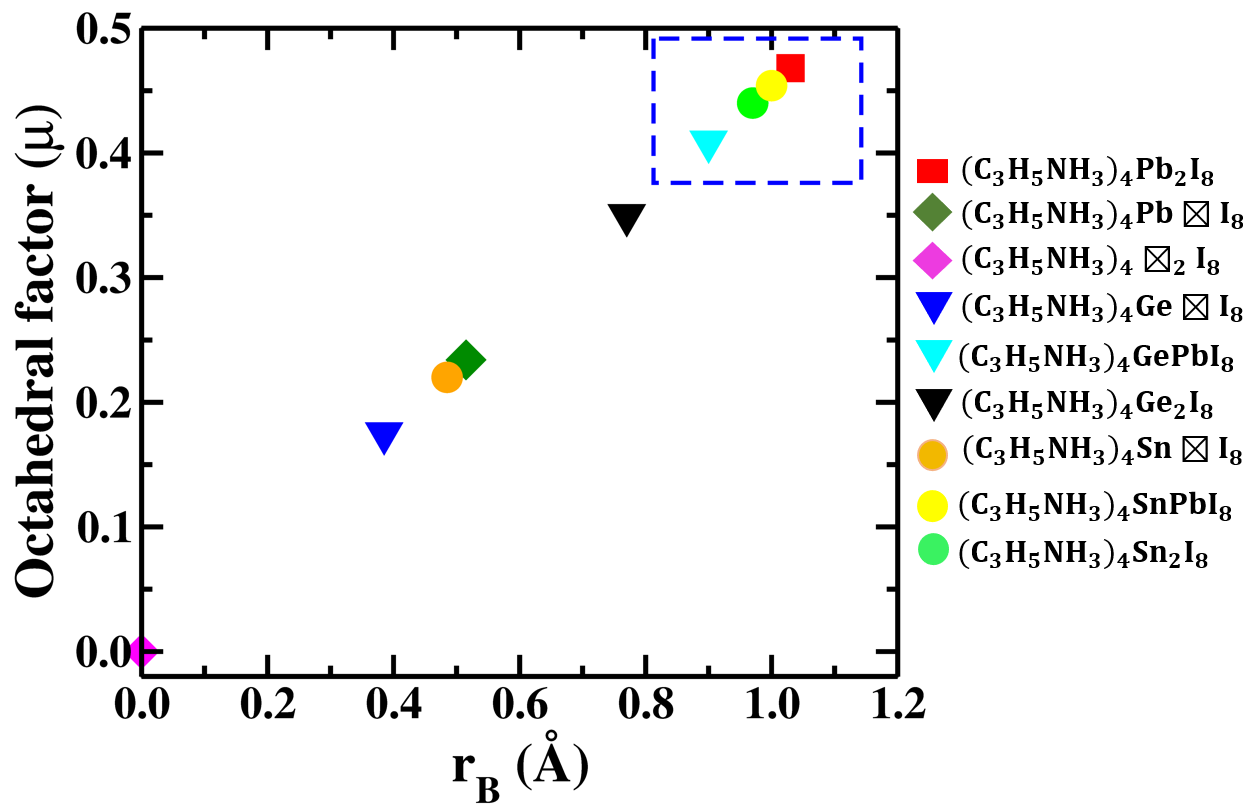}
  \caption{Calculated octahedral factor for different conformers.}
  \label{fig6}
\end{figure}

To investigate the structural stability of CPPI and all mixed conformers thoroughly, we have calculated Goldschmidt tolerance factor ($\emph t$),\cite{goldschmidt1926gesetze} of all the therodynamically stable configurations. This Goldschmidt tolerance factor indicates the structural stability of the perovskite structure, as defined in Equation \ref{eq:1}. We have found that all mixed conformers have $\emph t$ > 1, and form 2D perovskites.\cite{billing2007inorganic} However, the Goldschmidt tolerance factor alone is not sufficient to predict the stable perovskite structure. An additional criterion i.e., the octahedral factor ($\mu$) is considered, which determines the stability of the BX$_{6}$ octahedra,\cite{sun2017thermodynamic} defined as
\begin{equation}
\mu = \frac{r_\textrm{B}}{r_\textrm{X}}
\end{equation}
For stable BX$_{6}$ octahedra, the range of $\mu$ is 0.377 < $\mu$ < 0.895.\cite{sun2017thermodynamic} If the value of $\mu$ is beyond this range, then the perovskite structure will become unstable, even though the tolerance factor is in some favorable range for the perovskite formation.
The effective ionic radii of Pb$^{2+}$, Ge$^{2+}$, Sn$^{2+}$, and I$^{-}$ are 1.03, 0.77, 0.97, and 2.20 \AA, respectively.\cite{becker2017formation, travis2016application} The octahedral factor for all mixed conformers are shown in Fig. \ref{fig6} and given in tabular form in ESI† (Table S2). We have found that all mixed conformers with Pb-$\boxtimes$ and (C$_{3}$H$_{5}$NH$_{3}$)$_{4}$Ge$_{2}$I$_{8}$ have octahedral factor $\mu$ < 0.377 and do not possess the octahedral stability (see Fig. \ref{fig6}). Therefore, these are unstable perovskite structures, even though they have tolerance factor $\emph t $ > 1.0 and a favorable thermodynamic stability for 2D layered perovskite structures. The remaining mixed conformers, which are inside the blue box in Fig. \ref{fig6} have octahedral factor between the range 0.377 < $\mu$ < 0.895, and thus, these are structurally stable.

\subsection{Electronic Properties}
\begin{figure}[bh!]
  \centering
  \includegraphics[width=0.4\textwidth]{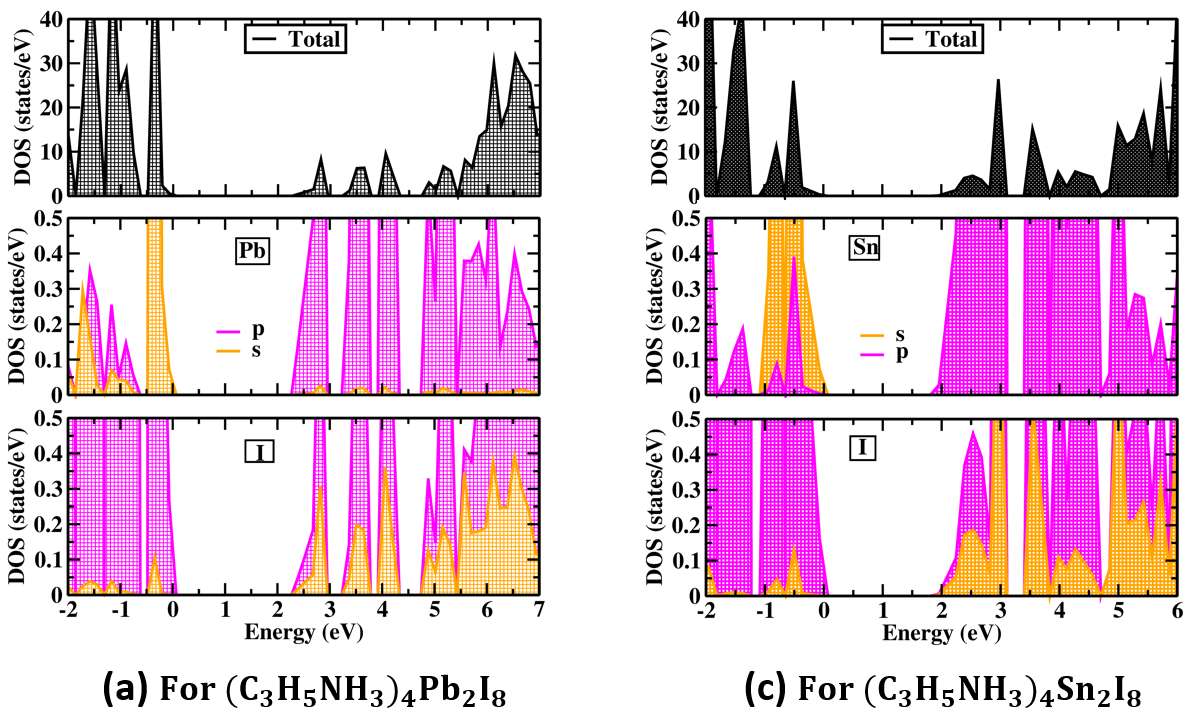}
  \includegraphics[width=0.4\textwidth]{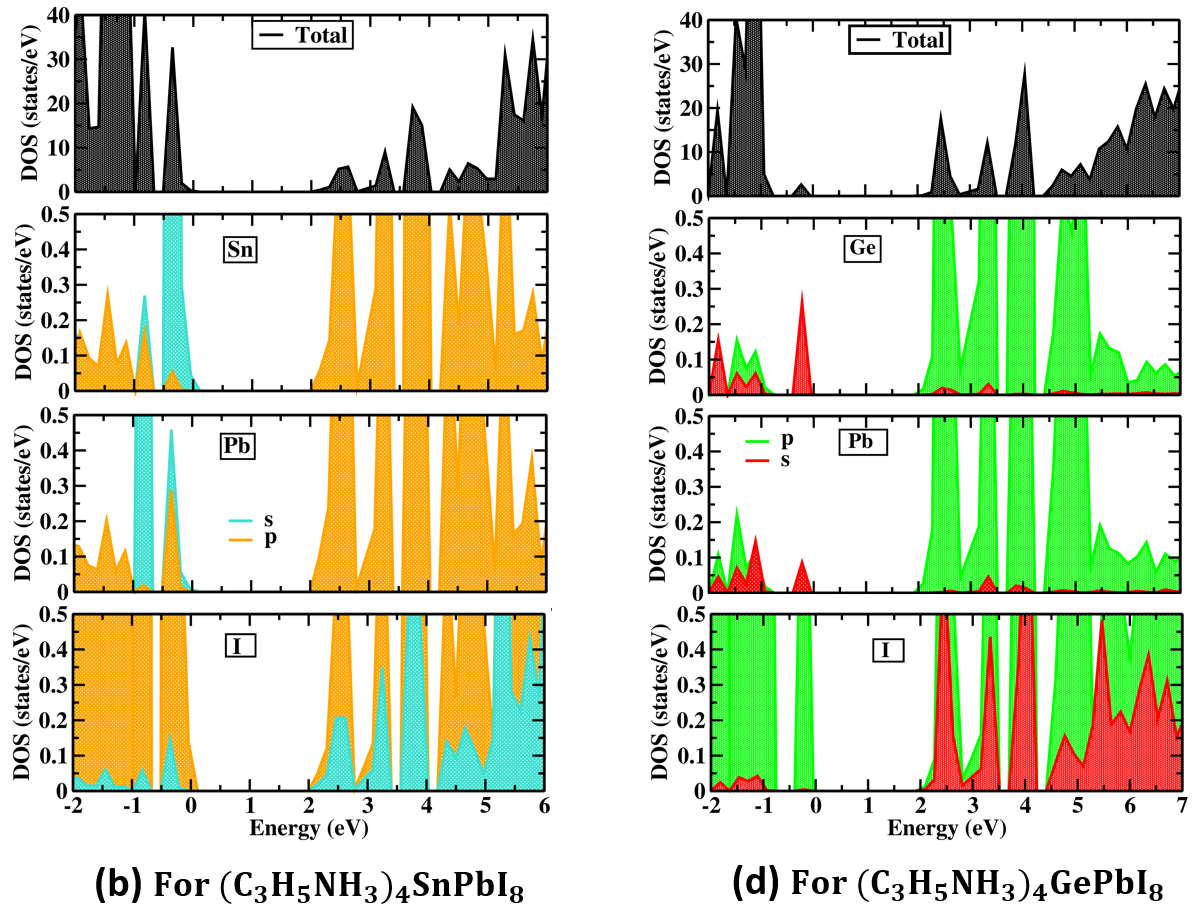}
  \caption{Calculated total and partial density of states for (a) (C$_3$H$_5$NH$_3$)$_4$Pb$_2$I$_8$, (b) (C$_{3}$H$_{5}$NH$_{3}$)$_{4}$SnPbI$_{8}$, (c) (C$_{3}$H$_{5}$NH$_{3}$)$_{4}$Sn$_{2}$I$_{8}$, and (d) (C$_{3}$H$_{5}$NH$_{3}$)$_{4}$GePbI$_{8}$ using the HSE06+SOC with $\alpha$ = 25\%. The VBM is set to 0 eV.}
  \label{fig7}
\end{figure}

We have observed that the mixed conformers with Pb-$\boxtimes$ have indirect bandgap (see upper panel of Fig. \ref{fig5}) and thus, poor optical performance. Therefore, the mixed conformers containing Pb-$\boxtimes$ are not suitable for solar cell applications. Hence, we have studied bandgap engineering by Ge and Sn substitutions only (i.e. without Pb-$\boxtimes$) (see Table \ref{tbl:example1}), where both VB and CB are shifted downward in the case of Sn substitution and upward in the case of Ge substitution. We have plotted pDOS for stable mixed conformers to understand the high optical absorption of IOHPs. Fig. \ref{fig7}(a) shows that Pb 6s and I 5p orbitals are mainly contributing to the VBM, whereas the CBm is mainly composed of unoccupied Pb 6p, I 5s, and I 5p orbitals. A similar orbitals contribution has been observed in the case of (C$_{3}$H$_{5}$NH$_{3}$)$_{4}$Sn$_{2}$I$_{8}$, where Sn 5s and I 5s orbitals are mainly contributing to the VBM with a small contribution by Sn 5p orbitals, whereas the CBm is composed of unoccupied Sn 5p, I 5s, and I 5p orbitals (see Fig. \ref{fig7}(c)). In the case of (C$_{3}$H$_{5}$NH$_{3}$)$_{4}$SnPbI$_{8}$, the VBM is primarily contributed by Sn 5s, Pb 6s, and I 5p orbitals with a small contribution by Pb 6p orbitals, whereas the CBm is mainly composed of unoccupied Sn 5p, Pb 6p, and I 5p orbitals with a small contribution by I 5s orbitals (see Fig. \ref{fig7}(b)). Similarly, we can observe the behavior of orbitals contribution in the case of (C$_{3}$H$_{5}$NH$_{3}$)$_{4}$GePbI$_{8}$, as shown in Fig. \ref{fig7}(d). Thus, a strong s–p and p–p coupling exist, that help in reducing the bandgap. Moreover, on increasing the Ge/Sn concentration, the bandgap is decreasing. Due to these direct p–p and s–p electronic transitions, strong absorption is expected in (C$_{3}$H$_{5}$NH$_{3}$)$_{4}$GePbI$_{8}$, (C$_{3}$H$_{5}$NH$_{3}$)$_{4}$SnPbI$_{8}$, and (C$_{3}$H$_{5}$NH$_{3}$)$_{4}$Sn$_{2}$I$_{8}$  materials~\cite{dehury2021structure}. Here, Sn, Ge and Pb show a similar contribution to the pDOS because of their similar valence electronic configurations. Thus, these electronic structure studies motivate us to explore the optical properties and theoretical maximum efficiency of the stable mixed conformers.
\begin{table}[bh!]
\small
  \caption{\ The bandgaps (E$_{g}$) calculated by HSE06+SOC with $\alpha$ = 25\% of mixed conformers}
  \label{tbl:example1}
  \begin{tabular*}{0.48\textwidth}{@{\extracolsep{\fill}}llll}
    \hline
   Conformers & E$_{g}$ (eV)& VB shift (eV)& CB shift (eV) \\
    \hline
    (C$_{3}$H$_{5}$NH$_{3}$)$_{4}$Pb$_{2}$I$_{8}$ &  2.30 &0.000 & 0.000 \\
    (C$_{3}$H$_{5}$NH$_{3}$)$_{4}$GePbI$_{8}$ & 2.13&$-$0.051 & $-$0.236 \\
    (C$_{3}$H$_{5}$NH$_{3}$)$_{4}$Ge$_{2}$I$_{8}$& 2.04 & $-$0.183&$-$0.447  \\ 
    (C$_{3}$H$_{5}$NH$_{3}$)$_{4}$SnPbI$_{8}$ & 2.04 & +0.348& +0.091\\
    (C$_{3}$H$_{5}$NH$_{3}$)$_{4}$Sn$_{2}$I$_{8}$ &1.94 & +0.583& +0.218 \\
    \hline
  \end{tabular*}
\end{table}
\subsection{Optical properties}
We can predict the suitability of the materials for optoelectronic applications by studying their optical properties such as dielectric function, extinction coefficient, refractive index and absorption coefficient. The linear optical properties are described by the frequency dependent complex dielectric function, $\varepsilon$($\omega$) = Im($\varepsilon$) + Re($\varepsilon$). Here, Im($\varepsilon$) and Re($\varepsilon$) are the imaginary and real part of the dielectric function, respectively. Using these, we have determined various optical properties, viz., refractive index ($\eta$), extinction coefficient ($\kappa$) and absorption coefficient ($\alpha$). The respective formulae are\cite{C9TC05002G,doi:10.1063/5.0031336}
\begin{equation}
\eta= \frac{1}{\sqrt{2}}\left[\sqrt{\textrm{Re}(\varepsilon)^2 + \textrm{Im}(\varepsilon)^2} + \textrm{Re}(\varepsilon)\right]^{\frac{1}{2}}
\label{eq:4}\\
\end{equation}
\begin{equation}
\kappa= \frac{1}{\sqrt{2}}\left[\sqrt{\textrm{Re}(\varepsilon)^2 + \textrm{Im}(\varepsilon)^2} - \textrm{Re}(\varepsilon)\right]^{\frac{1}{2}}
\label{eq:5}\\
\end{equation}
\begin{equation}
\alpha = \frac{2\omega\kappa}{c}
\label{eq:6}
\end{equation} 
where, $\omega$ is the frequency and $\emph c$ is the speed of light. The calculation of these optical properties is important for optoelectronic devices because these provide the response of the materials to incident electromagnetic radiations and demonstrate about the optimal solar energy conversion efficiency. 

\begin{figure}[tb!]
\centering
\includegraphics[width=0.4\textwidth]{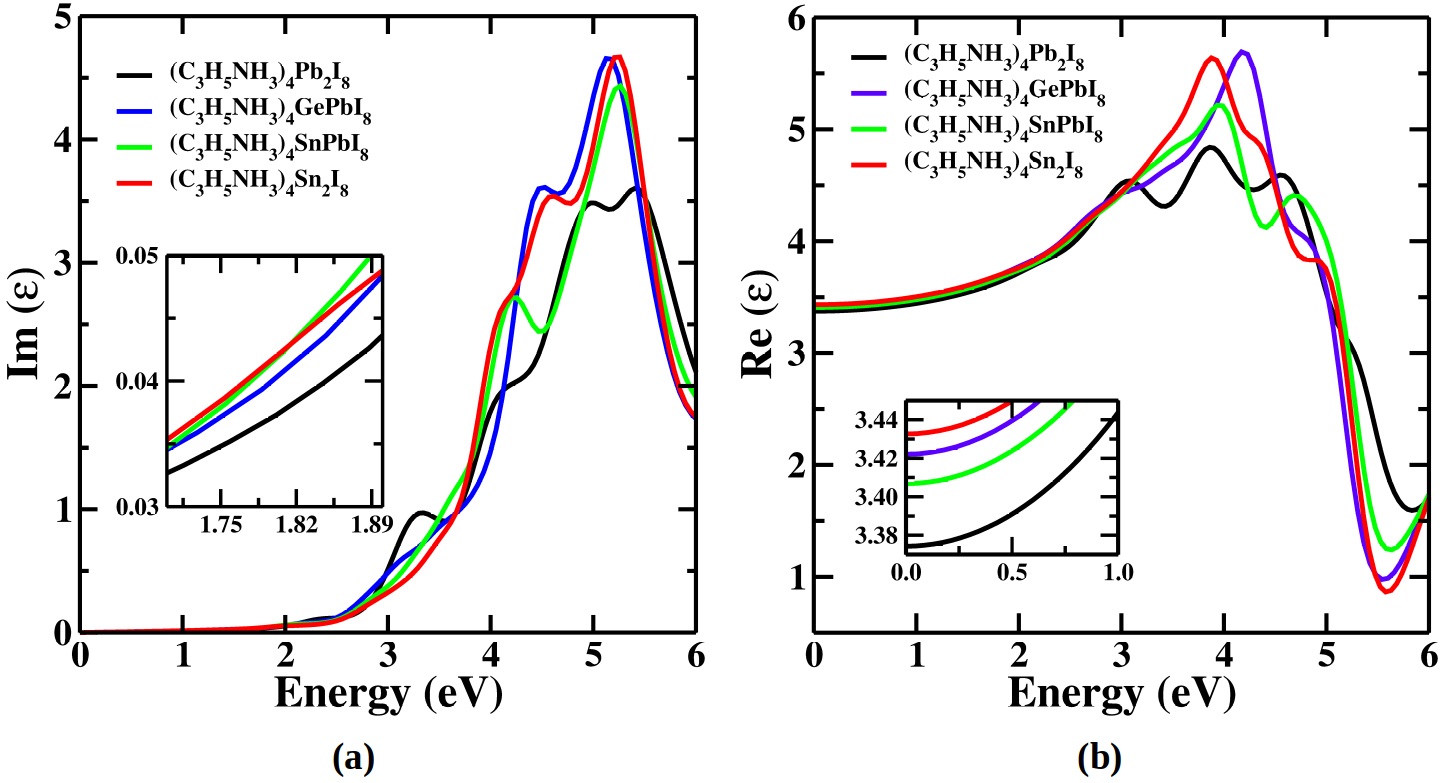}
\caption{\bf{(a)} {\normalfont Calculated imaginary part of the dielectric function, Im($\varepsilon$), and} \bf{(b)} {\normalfont calculated real part of the dielectric function, Re($\varepsilon$) for different stable conformers using HSE06+SOC with $\alpha$ = 25\%.}}
\label{fig8}
\vspace*{0.4cm}
\includegraphics[width=0.4\textwidth]{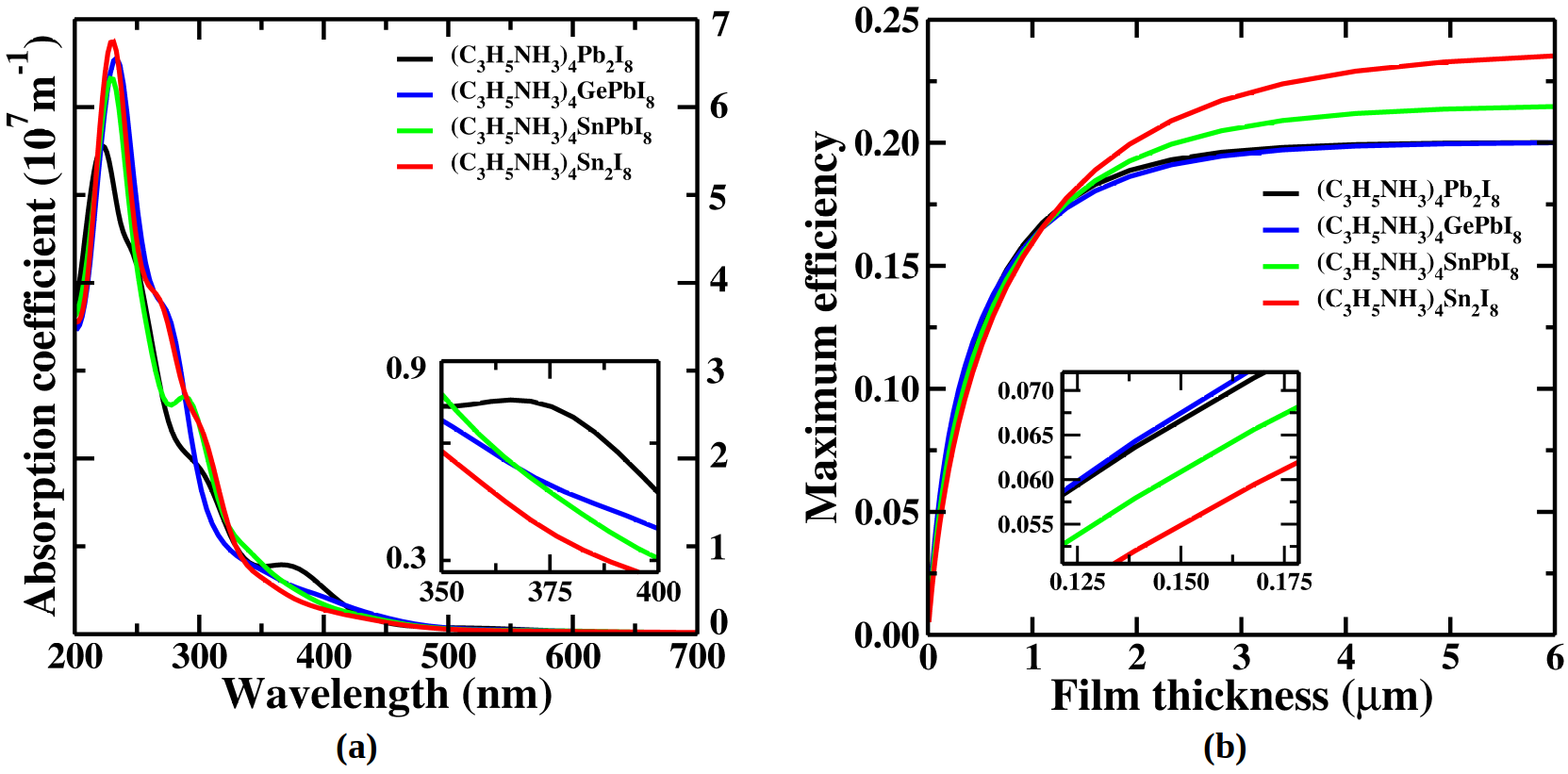}
\caption{\bf{(a)} {\normalfont Calculated absorption coefficient, and} \bf{(b)} {\normalfont SLME vs film thickness, of different stable conformers using HSE06+SOC with $\alpha$ = 25\%.}}
\label{fig9}
\end{figure}

Since the optical calculation is hugely dependent on the bandgap, if we consider SOC effect with HSE06 $\epsilon_{xc}$ functional, then the optical properties get underestimated because HSE06+SOC with $\alpha$ = 25\% hugely underestimates the CPPI bandgap (E$_\textrm{g}^\textrm{cal}$ = 2.30 eV, see Fig. \ref{fig4}(a)). To avoid this problem, we have calculated the optical properties by using both HSE06 and HSE06+SOC with $\alpha$ = 25\% and compared the results. The calculated imaginary and real part of the dielectric function, and the absorption coefficient for different stable conformers using HSE06 $\epsilon_{xc}$ functional are given in the ESI† (see Fig. S3 and S4). We have found that the lead-free mixed conformers follow the same trend using both HSE06 and HSE06+SOC. This is an expected result because those conformers do not contain the heavy element Pb, and thus, the inclusion/exclusion of SOC with HSE06 $\epsilon_{xc}$ functional has negligible effect on the results. The imaginary part of the dielectric function shows a red-shift towards the infrared region with increasing concentration of Sn/Ge (see Fig. \ref{fig8}(a), and a much clear view can be seen in Fig. S3(a) in ESI†). This is attributed to a decrement in the bandgap on increasing the amount of Sn/Ge. A large static value of the dielectric constant, i.e., Re($\epsilon$) (at $\omega$ = 0) is an essential requirement for an efficient solar absorber because it results in a high degree of charge screening, which can prohibit radiative electron-hole recombination and improves performance of the devices. From Fig. \ref{fig8}(b), we have observed a rise in value of Re($\epsilon$) (at $\omega$ = 0) with increasing Sn/Ge concentration. Sn and Ge substituted conformers have higher optical absorption peak intensity and red-shifted peaks in comparison to pristine CPPI within the UV region (see Fig. \ref{fig9}(a)).

\subsection{Spectroscopic limited maximum efficiency (SLME)}
To identify efficient materials with high power conversion efficiency (PCE) for PV applications, we have calculated SLME\cite{yin2014unique,yu2013inverse,doi:10.1021/acs.jpclett.1c01034} of different stable mixed conformers. Another way to select the efficient materials for solar cell applications is the Shockley–Queisser (SQ) limit,\cite{ruhle2016tabulated} but SQ limit only considers the bandgap of materials and does not take care of material's properties such as absorber layer thickness and absorption coefficient. Therefore, to overcome these shortcomings of SQ limit, Yu and Zunger proposed the SLME approach, which considers the absorber layer thickness and absorption coefficient of the system. It also depends on nature of the bandgap (direct or indirect), and temperature. Thus, we have used this SLME metric and calculated the maximum theoretical efficiency for all the stable mixed perovskite structures using HSE06 (see Fig. S4 in the ESI†) and HSE06+SOC with $\alpha$ = 25\% (see Fig. \ref{fig9}(b)). We have also tabulated the SLME values at 6 $\mu$m absorber thickness in Table \ref{tbl:example2}. The temperature is set to 300 K.
We have found that the conformer (C$_{3}$H$_{5}$NH$_{3}$)$_{4}$Sn$_{2}$I$_{8}$ has the maximum efficiency $\sim$ 24\%, which is higher than that of CPPI PCE (see Table \ref{tbl:example2}). Thus, we can conclude that complete removal of toxic element Pb with alternative Sn is possible with enhanced PCE. Therefore, we can recommend the substitution of Sn to replace toxic Pb completely, and to enhance the efficiency of solar cell based on 2D layered perovskites.
 \begin{table}[bh!]
\small
  \caption{\ SLME of different stable conformers at 6 $\mu$m absorber layer thickness}
  \label{tbl:example2}
  \begin{tabular*}{0.48\textwidth}{@{\extracolsep{\fill}}lll}
    \hline
   Conformers & SLME (HSE06) (\%) & SLME (HSE06+SOC) (\%)\\
    \hline
    (C$_{3}$H$_{5}$NH$_{3}$)$_{4}$Pb$_{2}$I$_{8}$ & 4.26 &20.02 \\
    (C$_{3}$H$_{5}$NH$_{3}$)$_{4}$GePbI$_{8}$ &11.16 & 20.02\\
    (C$_{3}$H$_{5}$NH$_{3}$)$_{4}$SnPbI$_{8}$ & 12.10 & 21.54\\
    (C$_{3}$H$_{5}$NH$_{3}$)$_{4}$Sn$_{2}$I$_{8}$& 23.85& 23.78\\
    \hline
  \end{tabular*}
\end{table}
\subsection{Wannier-Mott approach and exciton binding energy}
For a screened interacting electron-hole (e-h) pair the exciton binding energy (E$_\textrm{B}$) can be calculated employing Wannier-Mott approach~\cite{waters2020semiclassical}.  According to Wannier-Mott model E$_\textrm{B}$ for a system is defined as follows:
\begin{equation}
	\begin{split}
		\textrm{E}_\textrm{B}=\left(\frac{\mu}{\epsilon_\textrm{eff}^2}\right)\textrm{R}_\infty
		\label{eq9}
	\end{split}
\end{equation}
where, $\mu$, $\epsilon_\textrm{eff}$,  and R$_\infty$ are the reduced mass,  effective dielectric constant and Rydberg constant, respectively.  Note here that we have considered only electronic contribution to dielectric constant. Hence, for our case $\epsilon_\textrm{eff}$ = $\varepsilon_{\infty}$, where $\varepsilon_{\infty}$ corresponds to the electronic static dielectric constant. The effective mass of electrons and holes have been calculated using Wannier-Mott approach by plotting E-$\textit{k}$ dispersion curve (see Fig. \ref{fig6}) for pristine (C$_{3}$H$_{5}$NH$_{3}$)$_{4}$Pb$_{2}$I$_{8}$ and different configurations obtained after doing substitution at Pb.  The parabolic fitting of the dispersion curves have been done to compute the effective mass of the electrons and holes.  The effective mass can be calculated using following equation:   
\begin{equation}
\begin{split}
\textrm{m}^*=\frac{\hbar^2}{\frac{d^2\textrm{E}(k)}{dk^2}}
\label{eq1}
\end{split}
\end{equation}
where m$^*$ is the effective mass, E($\textit{k}$) is the energy, $\textit{k}$ is the wave vector, and $\hbar$ is the reduced Plank's constant. 
The calculated effective mass,  reduced mass in terms of rest mass of electron (m$_0$) and exciton binding energy are tabulated in Table~\ref{Table1}. From Table~\ref{Table1}, we have observed that these compounds exhibit large exciton binding energy.  On substituting Sn at Pb the exciton binding energy increases whereas it decreases when we substitute Ge at Pb.  Large exciton binding energy shows that electron-hole pairs are strongly bouned in these materials than in conventional lead halide perovskites~\cite{basera2020reducing}.

\begin{table}[htbp]
	\caption{Effective mass of hole m$_{\textrm{h}}^\ast$, electron m$_{\textrm{e}}^\ast$, reduced mass $\mu$ in terms of rest mass of electron m$_0$,  static value of electronic dielectric constant ($\varepsilon_{\infty}$),  and exciton binding energy E$_{\textrm{B}}$ (eV). } 
	\begin{center}
		\begin{tabular}[c]{cccccc} \hline
    Conformers & m$_{\textrm{h}}^\ast$ & m$_{\textrm{e}}^\ast$ & $\mu$  &    $\varepsilon_{\infty}$  & E$_{\textrm{B}}$ (eV)\\
   \hline
    (C$_{3}$H$_{5}$NH$_{3}$)$_{4}$Pb$_{2}$I$_{8}$ & -0.48 & 0.28 & 0.68  &  3.37 &  0.81 \\
    (C$_{3}$H$_{5}$NH$_{3}$)$_{4}$GePbI$_{8}$ & -0.47 & 0.25  &   0.52 &  3.42  &  0.61 \\
   (C$_{3}$H$_{5}$NH$_{3}$)$_{4}$Ge$_{2}$I$_{8}$ & -0.48     &    0.23  &   0.43 & 3.48    & 0.48  \\ 
    (C$_{3}$H$_{5}$NH$_{3}$)$_{4}$SnPbI$_{8}$ &  -0.40   &   0.26   &    0.76   &  3.41     &   0.90 \\
    (C$_{3}$H$_{5}$NH$_{3}$)$_{4}$Sn$_{2}$I$_{8}$&  -0. 31   &    0.24   &   1.00  &   3.43      &  1.15  \\
    \hline
		\end{tabular}
		\label{Table1}
	\end{center}
\end{table}
\subsection{Electron-phonon coupling strength}
Electron-phonon coupling is an alluring paradox as it influences the physical/chemical properties of a material.  In polar semiconductors (e.g., lead halide perovskites), the charge carriers interact with macroscopic electric field generated by longitudinal optical (LO) phonons, known as the Fr\"{o}hlich interaction.  Hence,  we have also studied electron-phonon coupling in our prototypical system ((C$_{3}$H$_{5}$NH$_{3}$)$_{4}$Pb$_{2}$I$_{8}$) using mesoscopic model, viz., Fr\"{o}hlich’s polaron model.  Fr\"{o}hlich coupling strength can be measured in terms of a dimensionless Fr\"{o}hlich parameter~\cite{frohlich1954electrons} $\alpha$ that is given as 

\begin{equation}
	\begin{split}
		\alpha = \frac{1}{4\pi\epsilon_0}\frac{1}{2}\left(\frac{1}{\epsilon_\infty}-\frac{1}{\epsilon_\textrm{static}}\right)\frac{\textrm{e}^2}{\hbar\omega_{\textrm{LO}}}\left({\frac{2\textrm{m}^*\omega_{\textrm{LO}}}{\hbar}}\right)^{1/2}
		\label{eq:9}
	\end{split}
\end{equation} 
where $\epsilon_{\infty}$ and $\epsilon_{\textrm{static}}$ correspond to the electronic and ionic static dielectric constants, respectively.  $\textrm{m}^*$ is the effective mass. $\epsilon_0$ is the permittivity of free space. The characteristic frequency $\omega_{\textrm{LO}}$ can be calculated by taking the spectral average of all the possible infrared active optical phonon branches~\cite{hellwarth1999mobility}. 
The calculated characteristic frequency and  electron-phonon coupling constant for  pristine ((C$_{3}$H$_{5}$NH$_{3}$)$_{4}$Pb$_{2}$I$_{8}$) are 3013.04 cm$^{-1}$ and 0.67, respectively.  Note that, the electron-phonon coupling constant is relatively smaller than that of hybrid halide perovskites~\cite{frost2017calculating}. Hence,  electron-phonon coupling is expected to be smaller in Sn/Ge substituted configurations as well. 

\section{Conclusions}
We have systematically investigated the structural and optoelectronic properties of (un)def ected 2D hybrid (C$_{3}$H$_{5}$NH$_{3}$)$_{2}$PbI$_{4}$, using first principles calculations. The spin-orbit splitting of conduction band is noticeable, which leads to a decrement in the bandgap. Therefore, SOC effect has been duly considered in all the calculations to determine accurate optical properties of mixed conformers. The 2D perovskite material CPPI is a wide bandgap semiconductor with a poor absorption spectrum. We have tuned the bandgap of CPPI system by substituting less toxic alternatives Ge and Sn in place of toxic element Pb, and observed the enhancement in the optoelectronic properties of the system. Similarly, we can tune the bandgap and enhance the optoelectronic properties in the case of compounds CBPI, CPEPI, and CHXPI. We have observed that complete removal of toxic Pb from CPPI is possible using Sn, whereas only partial replacement of Pb is possible with Ge. Moreover, the mixed conformers with Sn are more stable and have higher PCE in comparison to the conformers with Ge. Thus, we conclude that Sn substitution is more favorable in comparison to Ge substitution to replace toxic lead from CPPI. Lead-free 2D halide perovskite (C$_{3}$H$_{5}$NH$_{3}$)$_{2}$SnI$_{4}$ has highest efficiency with enhanced stability, which is useful for PV devices.  Pristine and mixed configurations exhibit large exciton binding energy.  The electron-phonon coupling is smaller than conventional lead halide perovskites.  These results give more promise for experimental realization of more these type of new lead-free 2D perovskites for optoelectronic devices.
\begin{acknowledgement}
D.G. acknowledges UGC, India, for the senior research fellowship [grant no. 1268/(CSIR-UGC NET JUNE 2018)]. GY acknowledges IIT Delhi for financial support. GY and DG thank Manish Kumar for helpful discussions. SB acknowledges financial support from SERB under core research grant (grant no. CRG/2019/000647) to set up his High Performance Computing (HPC) facility ``Veena'' at IIT Delhi for computational resources.
\end{acknowledgement}
\begin{suppinfo}
Band gap using different functionals and formation energy of conformers; Octahedral factor of different conformers; optical properties and spectroscopic limited maximum efficiency using HSE06 functional. 
\end{suppinfo}

\providecommand*{\mcitethebibliography}{\thebibliography}
\csname @ifundefined\endcsname{endmcitethebibliography}
{\let\endmcitethebibliography\endthebibliography}{}

\end{document}